\documentclass[prl]{revtex4}
\usepackage{graphicx,amsmath,amssymb,natbib,changes}
\usepackage{hyperref,siunitx,bm,comment}

\newcommand{\E}{\mathbf{E}}

\begin{document}

\title{A probabilistic framework for the control of systems\\
with discrete states and stochastic excitation}
\author{Gianluca Meneghello}
\email[Corresponding author: ]{gianluca.meneghello@gmail.com}
\altaffiliation[current address: ]{Department of Earth, Atmospheric and
  Planetary Sciences, Massachusetts Institute of Technology, Cambridge,
Massachusetts, 02139, USA.}
\affiliation{Flow Control Lab, UC San Diego, La Jolla, CA 92093-0411, USA.}
\author{Paolo Luchini}
\affiliation{DIMEC, Universit\`a di Salerno, Via Ponte don Melillo, 84084 Fisciano (SA), Italy}
\author{Thomas Bewley}
\affiliation{Flow Control Lab, UC San Diego, La Jolla, CA 92093-0411, USA.}

\date{\today} 


\begin{abstract}
  A probabilistic framework is proposed for the optimization of efficient
  switched control strategies for physical systems dominated by stochastic
  excitation.  In this framework, the equation for the state trajectory is
  replaced with an equivalent equation for its probability distribution
  function in the constrained optimization setting.  This allows for a
  large class of control rules to be considered, including hysteresis and a
  mix of continuous and discrete random variables.  The problem of steering
  atmospheric balloons within a stratified flowfield is a
  motivating application; the same approach can be extended to a variety of
  mixed-variable stochastic systems and to new classes of control rules.
\end{abstract}

\maketitle

\section{Introduction}

Optimal control theory is concerned with minimizing the energy required to
maintain a feasible phase-space trajectory within a fixed time-average measure
from a target trajectory \cite{lewis1995optimal}. This may be achieved by solving
the constrained optimization problem \cite{nocedal2006numerical}

\begin{subequations}\label{eqn:JPhaseSpace}
  \begin{align} 
    \label{eqn:JPhaseSpace1}
    \min_u\; &  |{\bm u}|_Q \\
    \label{eqn:JPhaseSpace2}
    \text{with }& \left\{
      \begin{matrix}
	|{\bm x}-\bar{{\bm x}}|_R = \text{constant} \\[0.8em]
	\dot{{\bm x}} = {\bm f} ({\bm x},{\bm u})+{\bm \xi},
      \end{matrix}
    \right. 
  \end{align}
\end{subequations}
where $\bm{u}(\bm{x})$ is a given feedback control rule, $\bm{x} = \bm{x}(t)$ and
$\bar{\bm{x}} = \bar{\bm{x}}(t)$ are the actual and target trajectories in
phase space, and the additive noise term $\bm{\xi}$ models the unknown or
uncertain components of the dynamical system. The norms $|\cdot|_Q$ and
$|\cdot|_R$ must be chosen to reflect the actual control
energy and the specific measure of interest of the system state, but are often limited
to $L_2$ or $L_\infty$ norms to make the optimization problem tractable. The present work is
motivated by the general inability of the formulation \eqref{eqn:JPhaseSpace} to treat
problems with mixed continuous and discrete random variables,
hysteretic behavior, and/or norms others than $L_2$ or $L_\infty$.

\begin{figure*}[t]
  \centering
  \includegraphics{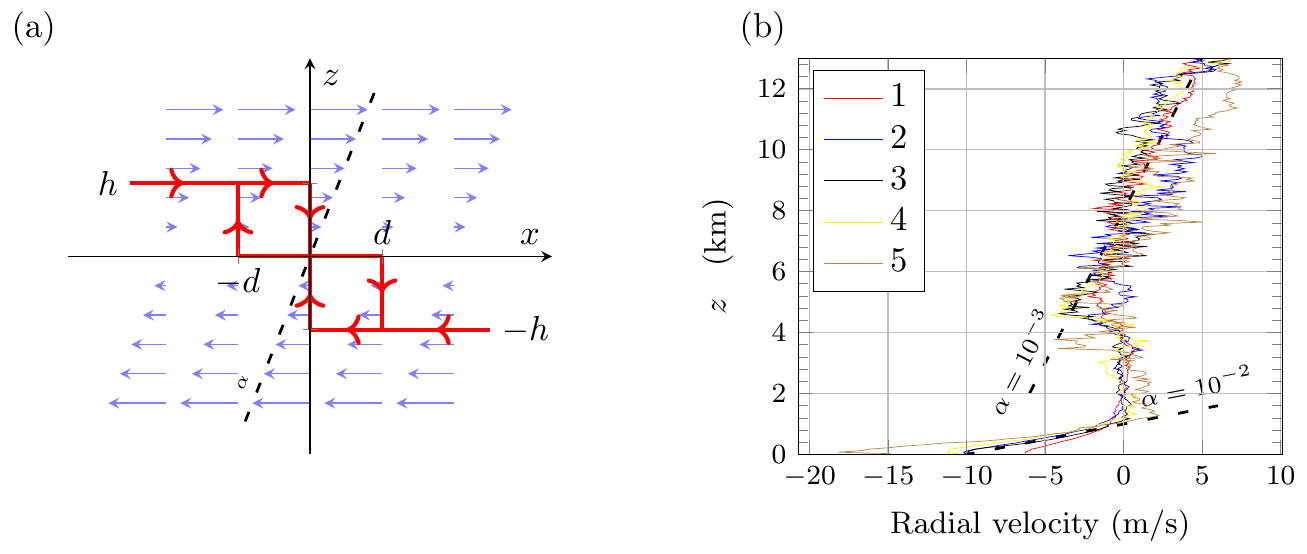}
  \caption{
    Left: flowfield model (thin arrows) and the three-level control (TLC)
    rule (thick lines). 
    Right: radial velocity profiles, composite from dropsonde measurements between 1996
    and 2012 within $\SI{200}{\kilo\meter}$ of the hurricane center \citep{wang2015long},
    binned into $\SI{50}{\meter}$ altitude intervals and sorted according to hurricane
    category (1 to 5).  The jaggedness of the profiles above \SI{2}{\kilo\meter}
    altitude is due to the reduced number of available measurements with
    respect to the lower region.
    Dashed lines estimate the mean velocity gradient $\alpha$.
  }
  \label{fig:flowField}
\end{figure*}

As a motivating application, consider a balloon in a stably stratified
turbulent flowfield whose time-averaged velocity is a function of height only,
as depicted by thin arrows in Figure~\ref{fig:flowField}a. This is
a good approximation for the radial flow within a hurricane, as depicted in
Figure~\ref{fig:flowField}b.
The balloon's density can be changed to control its vertical velocity (and,
hence, its altitude), and the balloon's motion can be well
approximated as the motion of a massless particle carried by the flowfield

\begin{subequations}   \label{eqn:system}
  \begin{align}
    \dot{X} &=  \alpha  Z + \xi, \\ 
    \dot{Z} &= u(X,Z),
  \end{align}
\end{subequations}
where $X$ and $Z$ are random variables denoting the horizontal and vertical
positions, $ \alpha$ is the vertical gradient of the time-averaged horizontal
velocity (i.e., $\alpha z$ is the time-averaged horizontal velocity at height
$z$), and the turbulent fluctuations of the horizontal velocities are
characterized by a white Gaussian noise $\xi$ with zero mean and spectral
density $c^2$. Neglecting the vertical velocity fluctuations, the balloon
moves in the horizontal direction according to a Brownian motion with a
probability distribution function (PDF) $p_{X,Z}(x,z)$ with
horizontal mean $\mu_X(t) = \alpha z t$ and variance $\sigma_X^2(t) =
c^2t$. In the uncontrolled case, the variance of the balloon's horizontal
position grows linearly with time.  The vertical velocity $u$ can then be
used, leveraging the background flow stratification $\alpha$, to return the
balloon to its original position.

We are thus interested in designing a
control strategy $u(x,z)$ to limit the variance of the horizontal position of the balloon to a target value
$\bar{\sigma}^2_X$, while minimizing the control cost $|u(x,z)|_Q$. More
specifically, we consider a three-level control (TLC) feedback rule,
depicted by thick lines in Figure~\ref{fig:flowField}a, consisting of
step-changes of altitude $\pm h$ in the vertical position, which are applied when the
balloon reaches a distance of $\mp d$ from the target trajectory $x=0$. In
such a setting, the vertical coordinate $\bar{Z} \in \left\{ -h , 0 , h
\right\}$ is discrete, and the control rule exhibits hysteresis in the horizontal
coordinate; we additionally note that the $L_1$ norm, measuring the step size
$h$, is a more appropriate measure of the energy $|\bm{u}|_Q$ required by the
balloon to change altitude than the classical $L_2$ norm. Despite the apparent
simplicity of this control rule, it cannot be optimized as
formulated in \eqref{eqn:JPhaseSpace}.

Rather than constraining the problem by the
state-space representation of the system \eqref{eqn:system}, as is done in
\eqref{eqn:JPhaseSpace}, we thus instead use an equivalent condition on the PDF
$p_{X,Z}(x,z)$, and 
restate the optimization problem \eqref{eqn:JPhaseSpace} as 

\begin{subequations}\label{eqn:JStochastic}
  \begin{align}
    \label{eqn:JStochastic1}
    \min_u \; &\E\left[ \,|\bm{u}|\, \right]\\
    \label{eqn:JStochastic2}
    \text{with } & \left\{
      \begin{matrix}
	\E\left[ \left(\bm{X-\bar{X}}  \right)^2 \right]  =
	\text{constant}\\[.8em]
	\partial_t p_{\bm{X}} + \bm{f}(\bm{X},\bm{u})\cdot \nabla
	p_{\bm{X}} + \dfrac{c^2}{2} \nabla^2 p_{\bm{X}} = 0,
      \end{matrix}
    \right. 
  \end{align}
\end{subequations}
where we have replaced the state equation in \eqref{eqn:JPhaseSpace2} with an
equivalent Fokker-Plank equation for the PDF
$p_{\bm{X}}(\bm{x})$ \citep{risken1984fokker}, and the norms are interpreted as
expected values. 
The solution of the optimization problem as stated in \eqref{eqn:JStochastic} is the principal contribution of this work.

The remainder of this paper is concerned with the solution of the optimization problem
\eqref{eqn:JStochastic} for the TLC rule of
Figure~\ref{fig:flowField}a, and with comparison to the classical
linear control rule $u = k_1x + k_2z$, whose optimal solution is given by the
Linear Quadratic Regulator (LQR) \cite{lewis1995optimal}. To facilitate comparison, we
first derive the functional form of the solution by dimensional analysis.

\section{Dimensional analysis}
\label{sec:dimensionalAnalysis}
The control problem is governed by three parameters: the velocity gradient
$ \alpha $, the spectral density $c^2$, and the target horizontal variance 
$\bar{\sigma}^2_X$. Take the length, time, and velocity scales as
$L=\sqrt{c^2/ \alpha }$, $T=\alpha^{-1}$, and $U=L/T=\sqrt{c^2 \alpha}$. A single
dimensionless parameter can be defined as
\begin{equation} 
  R = {\bar{\sigma}_X^2  \alpha }/{c^2},
  \label{eqn:Rdefinition}
\end{equation} 
and the dimensionless control cost can be written as
$w/U~=\E\left[\,|u|\,\right]/U~=~\mathcal{F}(R)$
where $\mathcal{F}(R)$ is an unknown dimensionless function. Similar
expressions can be written for $d/L$ and $h/L$. 
The system \eqref{eqn:system} is additionally invariant with respect to a
rescaling of the vertical coordinate by the time scale $ \alpha ^{-1}$, thus reducing
the parameters governing the problem to the standard deviation
$\bar{\sigma}_X$ and the spectral density $c^2$. Dimensional analysis
\citep{barenblatt1996scaling} can then be used to write the control cost as
\begin{equation}
  \label{eqn:dimensionalAnalysisCost}
  \dfrac{w}{U}      = \gamma_w \dfrac{1}{U}\dfrac{c^4}{ \alpha \sigma_X^3}  =
  \gamma_w \, R^{-\frac{3}{2}}.
\end{equation}
Expressions for the control parameters can also be obtained as
\begin{equation}
  \label{eqn:dimensionalAnalysisSteps}
  \dfrac{d}{L}  = \gamma_d \dfrac{\sigma_X}{L}  		 = \gamma_d R^{\frac{1}{2}}, \quad
  \dfrac{Z}{L}  = \gamma_h \dfrac{1}{L} \dfrac{c^2}{ \alpha \sigma_X} = \gamma_h R^{-\frac{1}{2}}.
\end{equation}

The solution is then summarized by the optimal value of $\gamma_{(\cdot)}$ for
each control parameter. Similarly, for the linear feedback control rule
$u=k_1x + k_2z$, we
can write
$Tk_1   = \gamma_{k_1} R^{-2} $  
and 
$Tk_2    = \gamma_{k_2} R^{-1}$. 
Note that the solution
\eqref{eqn:dimensionalAnalysisCost} is independent of the specific choice of
the control rule $u(x,z)$; a comparison between different rules can be
obtained by comparing the respective values of $\gamma_w$. 

\section{Three-level control (TLC) rule}
\label{sec:discreteControl}

\begin{figure}[h]
  ~\vskip-0.3in
  \centering
  \includegraphics{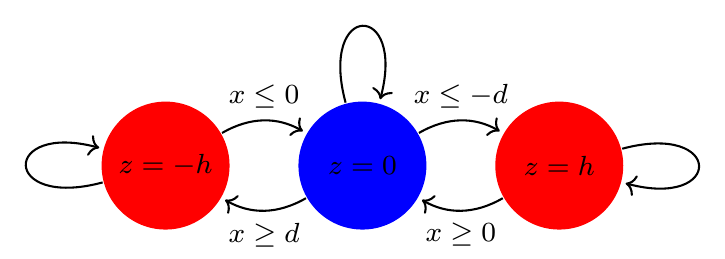}
  \caption{Implementation of the three-level control rule.}
  \label{fig:automata}
\end{figure}

We now proceed in seeking the optimal values for the parameters $d$ and $h$
[equivalently, $\gamma_d$ and $\gamma_h$ in
\eqref{eqn:dimensionalAnalysisSteps}] in the TLC rule
indicated by thick lines in Figure~\ref{fig:flowField}a, corresponding to
step changes in altitude $h$ at $x = 0,\pm d$.
In this limit, the governing equations \eqref{eqn:system} can be restated as 
\begin{equation}
  \dot{X} = - \alpha \bar{Z}+ \xi,
  \label{eqn:discreteSystem}
\end{equation}
where $X \in \mathcal{R}$ is the same as in the original problem, but $\bar{Z}
\in \left\{-h,0,h\right\}$ is now a discrete rather than a continuous random
variable, and the corresponding equation is replaced by the automata in
Figure~\ref{fig:automata}. This is a valid approximation when the control velocity $u$ is
larger than the horizontal velocity scale $\sqrt{c^2 \alpha }$.  Note that the dynamics of \eqref{eqn:discreteSystem} is quite simple,
and is dominated by the effect of the stochastic excitation by $\xi$.

Let $p_{X,\bar{Z}}\left( x,\bar{z} \right)$ be the PDF of the
balloon position, and 
$p_{\bar{z}}(x)=p_{X|\bar{Z}}\left(x|\bar{z}\right)\,p_{\bar{Z}}(\bar{z})$, so that 
$p_X(x)=\sum_{\bar{z}} p_{\bar{z}}(x)$ 
is the marginal probability. 
The governing equations for the PDFs can be written as

\begin{subequations}  \label{eqn:fokkerPlank}
  \begin{gather}
    \label{eqn:pZ}
    \begin{aligned}
      \partial_t p_{\bar{z}}(x)  +\partial_x  \alpha  \bar{z}\,  p_{\bar{z}}(x) -\partial_{xx} \dfrac{c^2}{2} p_{\bar{z}}(x)  = 0 & \\
      \textrm{for} \quad  \bar{z} \in \{\! &-\!h,0,h \},
    \end{aligned} \\
    \left.\partial_x p_X(x) \right|_{x^-} = \partial_x
    \left.p_X(x)\right|_{x^+} \quad \textrm{for}
    \quad  x \in \{ -d,0,d \},
    \label{eqn:transitionProbabilities}
  \end{gather}
\end{subequations}%
where \eqref{eqn:pZ} are three Fokker-Plank equations 
for each discrete altitude $\bar{z}$ obtained by considering the transition
probabilities implied by \eqref{eqn:discreteSystem}, and \eqref{eqn:transitionProbabilities} represent the
transition probabilities in the $z$ direction, and imposes the
conservation of the probability fluxes represented by arrows in
Figure~\ref{fig:automata}. Equations
\eqref{eqn:fokkerPlank} now take the role of the optimization problem
constraint \eqref{eqn:JStochastic2}.

\begin{figure}[ht]
  \centering
  \includegraphics{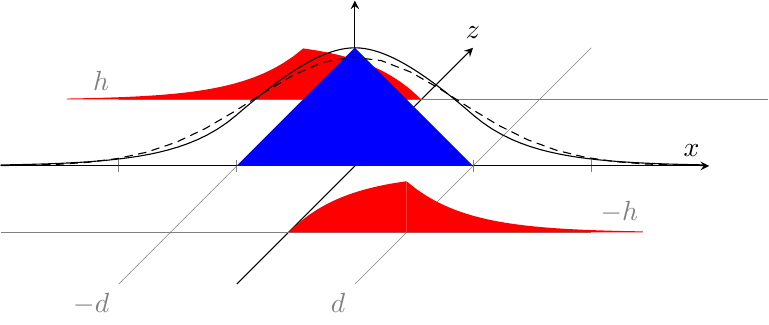}
  \caption{
    Steady state probability distribution functions for $c^2=h= \alpha
    =1$. Filled:
    $p_0(x)$ and $p_{\pm h}(x)$. Solid: marginal probability
    $p_X(x)=p_0(x)+p_{+h}(x)+p_{-h}(x)$. Dashed: normal
    distribution with the same variance.
  }
  \label{fig:steadyStatePDF}
\end{figure}
The statistically steady-state solution of \eqref{eqn:pZ} can actually be computed analytically, as shown in Figure
\ref{fig:steadyStatePDF}, and is given by

\begin{subequations}  \label{eqn:pdf}
  \begin{align}
    p_0\left( x\right) &= 
    \left\{
      \begin{array}{ll}
	c_1 x + c_2	& \qquad \qquad 0 < x < d,\\[1em]
	0			& \qquad \qquad d < x < \infty,
      \end{array}
    \right.  
    \\[1em]
    p_h\left( x\right) &=
    \left\{
      \begin{array}{ll}
	q_1 \frac{\lambda}{2} \; e^{-\frac{2x}{\lambda}} + q_2 & \quad 0 < x < d,\\[1em]
	r_1 \frac{\lambda}{2} \; e^{-\frac{2x}{\lambda}} + r_2 & \quad d < x < \infty, 
      \end{array}
    \right.  
  \end{align}
\end{subequations}
where $\lambda = \frac{c^2}{gh}$ is twice the e-folding scale of the PDF
(the symmetry of the problem at $x=0$ can be used to compute the solution for
$x<0$, $\bar{Z} = h$). The integration constants $c_1,c_2,q_1,q_2,r_1,r_2$ can
be obtained imposing \eqref{eqn:transitionProbabilities} together with the boundary conditions 
$p_0(x = d)                                          = 0  $, 
$p_h(x = 0)                                          = 0   $,
$\lim_{x\rightarrow \infty}  p_X(x)                    = 0   $,
and the normalization condition
$\int_{-\infty}^{\infty} p_X(x) dx                             = 1 $,
resulting in 

\begin{alignat}{3}
  c_1 &= - \dfrac{1}{d (d+\lambda)},\ \  &
  q_1 &= - \dfrac{1}{d (d+\lambda)},\ \  &
  r_1 &= \dfrac{e^{\frac{2d}{\lambda}}-1}{d(d+\lambda)},  \notag \\ 
  c_2 &= \dfrac{1}{d+\lambda}, &
  q_2 &= \dfrac{\lambda}{2 d (d+\lambda)}, &
  r_2 &= 0.
  \label{eqn:constants}
\end{alignat}
Upon substitution of the coefficients \eqref{eqn:constants} into the
expressions for $p_0$ and $p_h$ in \eqref{eqn:pdf}, the variance can be written
\begin{equation}
  \sigma_X^2 = \int_{-\infty}^{\infty} x^2 p_X(x)  dx = 
  \dfrac{d^3+2\lambda d^2+3\lambda^2 d+3 \lambda^3}{6(d+\lambda)}.
  \label{eqn:discreteVariance}
\end{equation}

The control cost can be computed as the total transition probability between
the states $\bar{Z} = 0$ and $\bar{Z} = h$, multiplied by the
cost of the single control activation $h$:
\begin{equation}
  w = 2  \alpha  h c^2 \left. \partial_x p_0\right|_{x=d} = 
  \dfrac{2 ghc^2}{d(d+\lambda)}.
  \label{eqn:discreteCost}
\end{equation}
The control parameters $h$ and $d$, and the corresponding control cost $w$,
can then be computed by minimization of the objective functional
\eqref{eqn:JStochastic1}.
The corresponding dimensionless constants
in \eqref{eqn:dimensionalAnalysisCost} and
\eqref{eqn:dimensionalAnalysisSteps} are

\begin{subequations}\label{eqn:discreteSolution}
  \begin{gather}
    \gamma_w = 0.5432,  \quad \gamma_d = 1.6288, \quad  \gamma_h = 1.1166, \\ f = 0.4864\, {c^2}/{\sigma_x^2} = 0.4864\, T^{-1}\,R^{-1},
  \end{gather}
\end{subequations}
where $f$ is the frequency at which the control has to be activated, and can
be computed by considering the times $t_{out} = d^2/c^2$ and
$t_{in} = d/(gZ)$ to reach the location $x=d$ from $x=0$ and back,
respectively.

\begin{figure*}[t]
  \centering
  \includegraphics{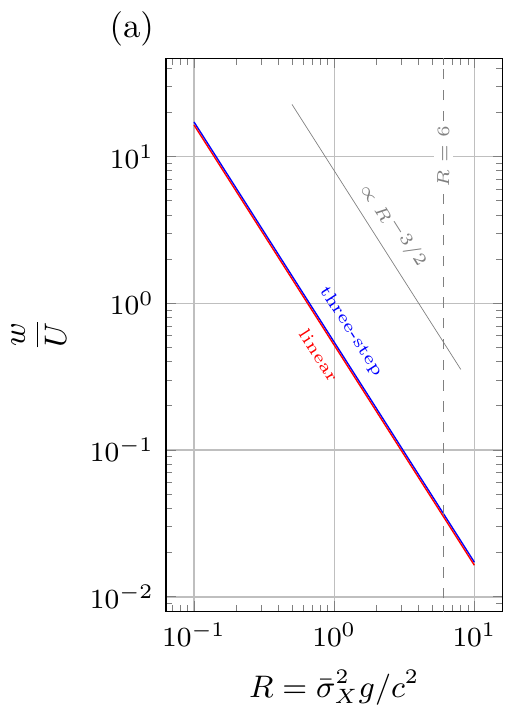}
  \includegraphics{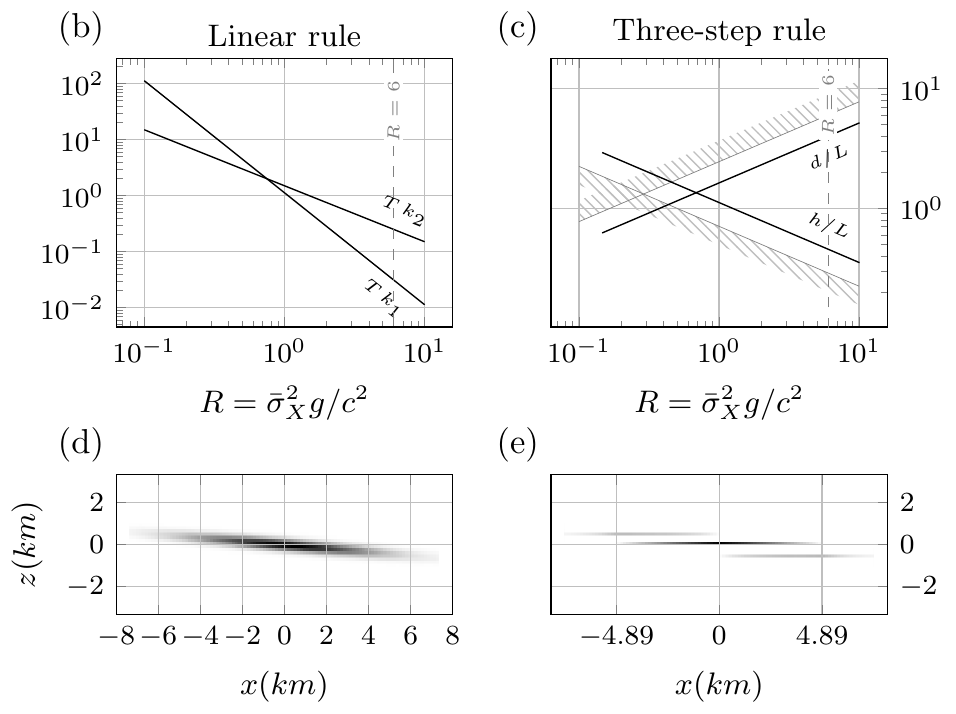}
  \caption{
    Comparison of control rules: (a) control cost $w=\E\left[ \,|u|\, \right]$
    as a function of the dimensionless parameter $R$ for the two different
    control rules; the two control costs are almost indistinguishable,
    differing by less than $5\%$.  (b,c) Control parameters and (d,e)
    corresponding probability density for the hurricane case $(R=6)$ for the
    linear and TLC control rules; gray patterns in (c) define the limit of
    each curve for which a given value of $R$ is attainable, given by
    \eqref{eqn:dhLimits}.
  }
  \label{fig:controlLawsResults}
\end{figure*}

It is also of interest to compute the minimum variance attainable
for given values of $d$ and $h$, and from there obtain the
limiting values of $d$ and $h$ for a specified $\bar{\sigma}_X$:

\begin{subequations}
  \label{eqn:dhLimits}
  \begin{align}
    \lim_{h\rightarrow \infty} \sigma_X^2 = \dfrac{d^2}{6} 
    &\quad \Rightarrow  \quad d < \sqrt{6}\,\bar{\sigma}_X,
    \\
    \lim_{d\rightarrow0} \sigma_X^2 = \dfrac{c^4}{2 \alpha ^2h^2}
    &\quad \Rightarrow  \quad h > \dfrac{1}{\sqrt{2}} \dfrac{c^2}{\bar{\sigma}_X  \alpha }.
  \end{align}
\end{subequations}
In both limits the control cost tends to infinity, in the first case because
$h\rightarrow\infty$, and in the second because the frequency of the steps increases without bound.
Reasonable (finite) values of $h$ and $d$, away from these limiting values, are thus important in application.
Results are summarized in Figure~\ref{fig:controlLawsResults}.

\section{Application to the control of balloons within a hurricane}
\label{sec:hurricaneControl}
Finally, we compute optimal values of the control parameters for atmospheric balloons within an
idealized hurricane flowfield, and compare them with results using the linear
control rule $u = k_1 x + k_2 z$. A velocity gradient of $\alpha =
\SI{e-3}{\per\second}$ (see Figure~\ref{fig:flowField}b) and a spectral
density of $c^2 = \SI{1500}{\meter\squared\per\second}$
\citep{zhang2011observational} are assumed.

We additionally impose a target standard deviation of $\bar{\sigma}_X =
\SI{3}{km}$, resulting in $R = 6$ [see \eqref{eqn:Rdefinition}]. Control
parameters and control cost can be obtained using
\eqref{eqn:dimensionalAnalysisSteps} together with the values in
\eqref{eqn:discreteSolution}. 
For the TLC rule,

\begin{subequations}
  \begin{align}
    \label{eqn:discreteResults}
    d &= \SI{4886.4}{m}, \quad
    &h &= \SI{558.3}{m},\\
    w &= \SI{4.54e-2}{m/s}, \quad
    &f &= \SI{8.11e-5}{s^{-1}},
  \end{align}
\end{subequations}
where $f$ corresponds to a period of about 3.5 hours.

The optimal solution for the \textit{linear} control rule can be readily
obtained by solving \eqref{eqn:JPhaseSpace}, resulting in

\begin{subequations}
  \begin{align}
    \label{eqn:continuousResults}
    k_1 &= \SI{3.125e-5}{s^{-1}},\quad
    &k_2 &= \SI{2.5e-4}{s^{-1}},\\
    w   &= \SI{4.32e-2} {m/s}.\qquad &&
  \end{align}
\end{subequations}
Simulations for both control rules are shown in
Figure~\ref{fig:controlLawsResults}d and \ref{fig:controlLawsResults}e.
\section{Conclusions}
\label{sec:conclusion}
This paper introduces a probabilistic framework for the optimization of
physical systems dominated by stochastic excitation in the presence of
mixed continuous and discrete random variables, non-linearities, and
hysteresis. Its application has been demonstrated by addressing the
problem of controlling
atmospheric balloons within a stratified flowfield; the same
framework can be
extended to a class of problems that, to the authors' knowledge, were
previously 
intractable from an optimization point of view. From an application
perspective, the three-level control (TLC) rule of Figure~\ref{fig:automata} has
many advantages: holding rather than continuously adjusting a position is
often an easier solution to implement, possibly requiring less energy in
real life applications.  For an observational platform like a sensor
balloon, it has the additional advantage of performing measurements which
are not disturbed by continuous control actions.


\appendix
\end{document}